\documentstyle[preprint,aps,fleqn]{revtex}

\begin{document}
\draft
\title{The mechanical relaxation study of polycrystalline MgC$_{1-x}$Ni$_3$}
\author{Y. X. Yao, X. N. Ying, Y. N. Huang$^{+}$, Y. N. Wang}
\address{National Laboratory of Solid State Microstructures and Physics Department,\\
Nanjing University, Nanjing 210093, China}
\author{Z. A. Ren, G. C. Che, H. H. Wen, and Z. X. Zhao}
\address{National Laboratory of Superconductivity, Institute of Physics, Chinese\\
Academy of Science, Beijing 100080, China}
\author{J.W. Ding}
\address{Department of Material Science and Engineering, University of Science and\\
Technology, Hefei 230026, China}
\maketitle

\begin{abstract}
The mechanical relaxation spectra of a superconducting and a
non-superconducting MgC$_{1-x}$Ni$_3$ samples were measured from liquid
nitrogen temperature to room temperature at frequency of kilohertz. There
are two internal friction peaks (at 300 K labeled as P1 and 125 K as P2) for
the superconducting sample. For the non-superconducting one, the position of
P1 shifts to 250 K, while P2 is almost completely depressed. It is found
that the peak position of P2 shifts towards higher temperature under higher
measuring frequency. The calculated activation energy is 0.13eV. We propose
an explanation relating P2 to the carbon atom jumping among the off-center
positions. And further we expect that the behaviors of carbon atoms maybe
correspond to the normal state crossovers around 150 K and 50 K observed by
many other experiments.

$^{+}$Author to whom correspondence should be addressed. Email address:
ynhuang@nju.edu.cn
\end{abstract}

\begin{center}
\end{center}

\newpage

The superconductivity has been found in intermetallic compounds of the
general formula MgC$_{1-x}$Ni$_3$ [1], which has the antiperovskite
structure. Many researches have been focused on the possible magnetism or
magnetic instability due to a large amount of Ni in this compound [2-5]. The
MgC$_{1-x}$Ni$_3$ band structure displays a characteristic very large and
narrow peak in the density of states (DOS) just below the Fermi energy
[6,7]. Then it is expected that the magnetic interaction still exists in MgC$%
_{1-x}$Ni$_3$ and there are magnetic fluctuations in superconducting MgC$%
_{1-x}$Ni$_3$ [8], which is reminiscent of the LnNi$_2$B$_2$C family [9].

Another aspect of MgC$_{1-x}$Ni$_3$ is the possibility of conventional
phonon-coupled paring [6,10]. The light element, carbon, might play an
important role on the superconductivity in MgC$_{1-x}$Ni$_3$ in the frame of
electron-phonon coupling. This viewpoint is also associated with the
recently found MgB$_2$ compound with the superconducting transition
temperature about 40 K [11]. The light element boron results in the high
phonon frequency, and hence the higher transition temperature.

There are some unresolved questions in the normal state such as the
crossover behavior located around 50 K and the ``knee'' near 150 K
[8,12,13]. It is also found that the superconductivity is critically
dependent on the carbon concentration in MgC$_{1-x}$Ni$_3$. When $1-x$ is
smaller than about 0.85, the superconductivity is completely suppressed and
the ``knee'' near 150 K disappears. While when $1-x$ is larger than 0.85
both the superconductivity and the ``knee'' recover. In this report, we
measured the mechanical relaxation spectra of a superconducting ($x\approx 0$%
) and a non-superconducting ($x\approx 0.2$) MgC$_{1-x}$Ni$_3$. The
comparative study would throw some insights into the mechanism for the
superconductivity of MgC$_{1-x}$Ni$_3$.

Two ceramic samples of MgC$_{1-x}$Ni$_3$ with different carbon content were
prepared by conventional method as described in Ref. 12. Resistance of the
samples was measured by the standard four-probe technique and the result was
shown in figure 1. The transition temperature T$_C$ of the superconducting
MgC$_{1-x}$Ni$_3$ sample is 7 K.

Internal friction ($Q^{-1}$) and modulus ($Y$) of the mechanical relaxation
spectra were measured by Frequency-Modulation-Acoustic-Attenuation (FMAA-I)
equipment made in the University of Science and Technology of China. The
dimensions of the rectangular bars of the samples are \symbol{126}5$\times $1%
$\times $0.2mm$^3$ with resonant frequencies from 2600 to 4100 Hz.

The mechanical relaxation spectra of a superconducting and a
non-superconducting samples on heating, with the rate about 1 K/min, were
shown in figure 2. For the superconducting sample, there are two internal
friction peaks (at 300 K labeled as P1 and 125 K as P2). With increasing the
measured frequency, the position of P2 shifts towards higher temperature,
which is shown in figure 3. Then it is expected that P2 may be due to the
thermally activated process. The calculated activation energy $U=0.13eV$ and
the pre-exponential factor $\nu _0=2.9\times 10^9$ Hz, by a fit of the
Arrhenius relation $2\pi f_p=\nu _0\exp (-U/T_p)$, where $f_p$ is the
resonant frequency and $T_p$ peak temperature. But P1 does not show clear
shift within experimental errors. For the non-superconducting sample, P2 is
nearly completely suppressed and P1 has shifted towards 250 K.

The relative changes of modulus of the two samples $\Delta Y/Y$, which can
be obtained by the following formula $\Delta Y/Y=(f^2-f_0^2)/f_0^2$, are
also shown in the figure 2, where $f$ is the resonant frequency, and $f_0$
the resonant frequency at 81 K. No clear softening point (local minimum) of
modulus was observed in our measured temperature range, which indicates
there is no lattice instability. It is consistent with the powder
diffraction experiment results [14], which shows that the lattice parameter
and the Debye-Waller factors for individual atoms decrease smoothly with
decreasing temperature, and no unusual change of the structure parameters
near T$_C$. This is different to the case for superconducting BKBO with the
perovskite structure, where structure instability was observed [15,16]. On
the other hand, there are modulus defects corresponding to P1 and P2 peaks
for the superconducting sample. But only one modulus defect appears around
260 for the non-superconducting one.

In figure 4, the prediction of Debye theory, $Q^{-1}=\frac \Delta T\frac{%
\varpi \tau }{1+\varpi ^2\tau ^2}$, is compared with P2 peak, where $\Delta $
is the relaxation strength, $\varpi =2\pi f$ , and $\tau =\nu _0^{-1}\exp
(U/T_p)$. It is clear that the prediction is much narrower than the
experiments. So, the Cole-Cole law [17], $Q^{-1}=\frac \Delta T%
\mathop{\rm Im}%
[\frac 1{1+(i\varpi \tau )^\alpha }]$, is used to fit the experimental
results. In the fitting, background internal friction, $Q_B^{-1}=4.5\times
10^{-6}T+7.6\times 10^{-4}$, that may originate from the low temperature
tail of P1 peak $etc$., is firstly subtracted. The fitted $\alpha =0.52$
that means there exist strong correlations between the relaxation units [18].

Due to both the small activation energy and small pre-exponential factor of
the relaxation time of P2 peak, $U=0.13eV$ and $\nu _0=2.9\times 10^9$ Hz,
we expect that it is related to the off-center hoping of carbon atoms [19].
Usually, the perovskite oxide presents the ferroelectricity, which is due to
the ordering of off-center light ions in the octahedral structure. In MgC$%
_{1-x}$Ni$_3$, light element carbon locates in the center of octahedron.
Then carbons might posses the off-center configurations and hop among the
off-center sites. It should be noted that this off-centered configuration
does not alternate the average crystalline symmetry and average symmetry of
carbon atoms when no ordering takes place. Then it is not easy to be
observed by other experiments. This kind of off-center hopping leads to
mechanical energy loss and modulus defect, and the activation energy and the
pre-exponential factor are usually small, such as the off-center relaxation
process of oxygen atoms in YBa$_2$Cu$_3$O$_{7-\delta }$ [20].

P2 peak is observed only in superconducting sample, and in
non-superconducting samples it is largely depressed. This might provide some
important information on the superconducting in MgC$_{1-x}$Ni$_3$. Ren $et$ $%
al.$ [12] investigated the structure feature of a series of MgC$_{1-x}$Ni$_3$
in detail. They observed two different phases of MgC$_{1-x}$Ni$_3$,
non-superconducting $\alpha $-MgC$_{1-x}$Ni$_3$ phase and superconducting $%
\beta $-MgC$_{1-x}$Ni$_3$ phase. Due to the smaller lattice parameter of $%
\alpha $-MgC$_{1-x}$Ni$_3$ phase, the off-center configurations would be
suppressed. It is consistent with the result that P2 is depressed in the
non-superconducting sample.

The off-center configurations will be smeared with the increase of
temperature and consequently, there will be a crossover point somewhat above
P2 peak temperature (\symbol{126}125 K). This might correspond to the
``knee'' near 150K observed in the resistivity, thermal power and thermal
conductivity in superconducting MgC$_{1-x}$Ni$_3$ [12,13]. Moreover, there
exist strong correlations among the off-center carbon atoms as mentioned
above, which will lead to their ordering at low temperature. Whether this
ordering is related to the electronic crossover near 50 K [8,13] needs
further studies in superconducting MgC$_{1-x}$Ni$_3$.

As to the P1 peak, due to its appearing in both superconducting and
non-superconducting samples, it is speculated that it may be related to the
jumping of carbon vacancies between different octahedrons. Usually, this
kind of jumping process is almost impossible in perovskite, but MgC$_{1-x}$Ni%
$_3$ is the so-called anti-perovskite, where nickel atoms, other than oxygen
ones, form the octahedrons, so that the binding to the center atoms is weak.
Moreover, carbon is light element that is easy to jump. Furthermore, the
nickel atoms nearest to the carbon vacancies shift from the idealized
positions [21], which will increase the mobility of carbon vacancies and
create local distortions. As a result, internal friction peak and modulus
defect is observed.

In summary, we have measured the mechanical relaxation spectra of
superconducting and non-superconducting MgC$_{1-x}$Ni$_3$ samples from
liquid nitrogen temperature to room temperature. A thermally activated
relaxation peak is observed around 125K in superconducting sample. We expect
that it should be due to the off-center hopping of carbon atoms. This
internal friction peak is almost fully depressed in non-superconducting
samples. This indicates that this process is closely related to the
superconductivity in MgC$_{1-x}$Ni$_3$. The other peak near 300K, which
appears in both superconducting and non-superconducting samples, is
speculated to be related to the jumping of carbon vacancies between
different octahedrons.

\section{Acknowledgments}

This work is supported by the National Science Foundation of China (Grant
No. 0204G141) and by the Ministry of Science and Technology of China
(NKBRSF-G19990646).

\begin{center}
Figure Captions
\end{center}

Figure 1. Reduced resistance $R/R_{300K}$ of a superconducting ($x\approx 0$%
) and a non-superconducting ($x\approx 0.2$) MgC$_{1-x}$Ni$_3$ samples vs
temperature $T(K)$.

Figure 2. Internal friction ($Q^{-1}$) and reduced modulus ($\Delta
Y/Y_{81K} $) of the mechanical relaxation spectra of a superconducting and a
non-superconducting MgC$_{1-x}$Ni$_3$ samples from liquid nitrogen
temperature to room temperature.

Figure 3. The internal friction spectra ($Q^{-1}$) of the P2 peak for the
superconducting MgC$_{1-x}$Ni$_3$ sample with the resonant frequency 2605
and 4098 Hz, respectively.

Figure 4. A fit of Cole-Cole law (solid line) to the P2 peak. The dashed
line is the prediction of the Debye theory.


\begin{references}
\bibitem{1}  T. He $et$ $al$, Nature 411, 54 (2001)

\bibitem{2}  C. M. Granada, C. M. da Silva, and A. A. Gomes, Solid State
Common. 122, 269 (2002)

\bibitem{3}  I. R. Shein $et$ $al$, Phys. Rev. B 66, 024520 (2002)

\bibitem{4}  In Gee Kim, Jae Il Lee, and A. J. Freeman, Phys. Rev. B 65,
064525 (2002)

\bibitem{5}  H. Rosner $et$ $al$, Phys. Rev. Lett. 88, 027001 (2002)

\bibitem{6}  J. H. Shim, S. K. Kwon, and B. I. Min, Phys. Rev. B 64, 180510
(2001)

\bibitem{7}  J. L. Wang $et$ $al$, J. Appl. Phys. 91, 8504 (2002)

\bibitem{8}  P. M. Singer $at$ $al$, Phys. Rev. Lett. 87, 257601 (2001)

\bibitem{9}  R. J. Cava $et$ $al$, nature 367, 252 (1994)

\bibitem{10}  S. B. Dugdale, and T. Jarlborg, Phys. Rev. B 64, 100508 (2001)

\bibitem{11}  J. Nagamatsu $et$ $al$, nature 410, 63 (2001)

\bibitem{12}  Z. A. Ren $et$ $al$, Physica C 371, 1 (2002)

\bibitem{13}  S. Y. Li $et$ $al$, Phys. Rev. B 65, 064534 (2002)

\bibitem{14}  Q. Huang $et$ $al$, Physica C 363, 215 (2001)

\bibitem{15}  S. Zherlitsyn $et$ $al$, Eur. Phys. J. B 16, 59 (2000)

\bibitem{16}  M. Braden $et$ $al$, Phys. Rev. B 62, 6708 (2000)

\bibitem{17}  K.S. Cole and R.H. Cole, J. Chem. Phys. 9, 341(1941)

\bibitem{18}  K.L. Ngai, Y.N. Wang and L.B. Magalass, J. Alloy \& Comp.
211/212, 327(1994)

\bibitem{19}  A.S. Nowick, Proc. of ICIFUAS-9, Pergamon Press, New York, ed.
by T.S. Ke, 85(1985)

\bibitem{20}  G. Cannelli $et$ $al$, Phys. Rev. B 42, 7925 (1990)

\bibitem{21}  T. G. Amos $et$ $al$, Solid State Common. 121, 73 (2002)
\end{references}
\end{document}